\documentclass[conference]{IEEEtran}
\IEEEoverridecommandlockouts

\usepackage[utf8]{inputenc}
\usepackage{cite}
\usepackage{amsmath,amssymb,amsfonts}
\usepackage{algorithmic}
\usepackage{subfig}
\usepackage{graphicx}
\usepackage{textcomp}
\usepackage{xcolor}
\usepackage{pifont}
\usepackage{multirow}
\usepackage{soul}
\usepackage{blkarray} 
\usepackage[ruled,vlined]{algorithm2e} 
\usepackage{capt-of}
\usepackage{caption}
\usepackage{array}
\usepackage{hyperref}

\usepackage[a4paper, total={184mm,239mm}]{geometry}
\def\BibTeX{{\rm B\kern-.05em{\sc i\kern-.025em b}\kern-.08em
T\kern-.1667em\lower.7ex\hbox{E}\kern-.125em}}

\usepackage[export]{adjustbox}%

\usepackage{tikz}

\newcommand{\circleNumber}[1]{%
  \begin{tikzpicture}[baseline=(char.base)]
    \node[shape=circle, fill=black, inner sep=2pt] (char) {\textcolor{white}{#1}};
  \end{tikzpicture}
}

\begin{document}

\title{CycPUF: Cyclic Physical Unclonable Function}

\author{
\IEEEauthorblockN{Michael Dominguez}
\IEEEauthorblockA{\textit{Computer Engineering \& Computer Science Department} \\
\textit{California State University Long Beach}\\
Long Beach, CA, USA \\
michael.dominguez01@student.csulb.edu} 
\and
\IEEEauthorblockN{Amin Rezaei}
\IEEEauthorblockA{\textit{Computer Engineering \& Computer Science Department} \\
\textit{California State University Long Beach}\\
Long Beach, CA, USA \\
amin.rezaei@csulb.edu} 
}

\maketitle

\begin{abstract}
Physical Unclonable Functions (PUFs) leverage manufacturing process imperfections that cause propagation delay discrepancies for the signals traveling along these paths. While PUFs can be used for device authentication and chip-specific key generation, strong PUFs have been shown to be vulnerable to machine learning modeling attacks. Although there is an impression that combinational circuits must be designed without any loops, cyclic combinational circuits have been shown to increase design security against hardware intellectual property theft. In this paper, we introduce feedback signals into traditional delay-based PUF designs such as arbiter PUF, ring oscillator PUF, and butterfly PUF to give them a wider range of possible output behaviors and thus an edge against modeling attacks. Based on our analysis, cyclic PUFs produce responses that can be binary, steady-state, oscillating, or pseudo-random under fixed challenges. The proposed cyclic PUFs are implemented in field programmable gate arrays, and their power and area overhead, in addition to functional metrics, are reported compared with their traditional counterparts. The security gain of the proposed cyclic PUFs is also shown against state-of-the-art attacks.
\end{abstract}


\section{Introduction}
\label{Introduction}
With an ever-increasing amount of sensitive data stored in electronic devices and the growing involvement of third-party manufacturing foundries, design automation companies, and testing facilities, hardware security is becoming an increasingly important issue in today's digital world. Physical Unclonable Functions (PUFs) \cite{Abbott2020} are hardware security primitives that ideally generate a unique device-dependent response to a particular input stimulus, such as a challenge or an interrogation signal. These responses are based on the inherent and uncontrollable physical variations that occur during the manufacturing process of Integrated Circuits (ICs). A weak PUF supports a small number of Challenge-Response Pairs (CRPs) with a polynomial function of the PUF size, while a strong PUF scales in such a way that it can support a very large number of CRPs with exponential growth with respect to the PUF size. In recent years, the use of PUFs in hardware security problems has gained popularity due to their ability to provide a unique identity for every chip \cite{Saqib2015}, generate unpredictable hardware-based keys \cite{Devadas2007, Homma2020}, and facilitate privacy-friendly authentication protocols \cite{Aysu2015}. PUFs have been extensively studied in academia and have found their way into commercial applications \cite{Intrinsic-ID2023, Secure-IC2023, Fungible2023}, making them a promising solution for hardware security problems such as watermarking \cite{Urien2021} and logic locking \cite{Rezaei2022, Rezaei:BreakUnroll, Maynard:DK-Lock, Aghamohammadi:CoLA}.

One of the main challenges in implementing PUFs is their vulnerability to modeling attacks \cite{Yang2021, Chakraborty2019, Kundu2016, Zwolinski2021}. In such attacks, adversaries usually use Machine Learning (ML) techniques to model the PUF based on a limited number of observed CRPs and predict its responses to new challenges, which poses a significant threat to the security of PUF-based systems. While researchers have developed hard-to-model PUF designs \cite{Chongyan2021, Lao2021, Karimi2021, Ruhrmair2018}, their complexity imposes a high design overhead and has a negative effect on PUF functional metrics such as uniqueness, uniformity, and reliability.

Contrary to conventional wisdom, which believes that combinational circuits must be free of feedback loops, cyclic combinational circuits have been proposed for high-speed and low-power designs \cite{Malik1994, Riedel2003}, and their application in logic locking has been studied \cite{Rezaei2018, Rezaei2019} to improve security against oracle-guided attacks \cite{Subramanyan2015}. Inspired by these seminal works, in this paper, we propose CycPUF, a \ul{Cyc}lic \ul{P}hysical \ul{U}nclonable \ul{F}unction framework, to safeguard against modeling attacks and enhance the PUF functional metrics with reasonable overhead. Our contributions are as follows:

\begin{itemize}
    \item [$\bullet$] Proposing three delay-based CycPUFs and thoroughly analyzing their output behavior;
    \item [$\bullet$] Comparing CycPUFs functional metrics and overhead with their acyclic counterparts;
    \item [$\bullet$] Showcasing the security gain of CycPUFs against ML-based and hybrid modeling attacks.
\end{itemize}

\begin{figure*}
    \centering
    \includegraphics[width=\textwidth]{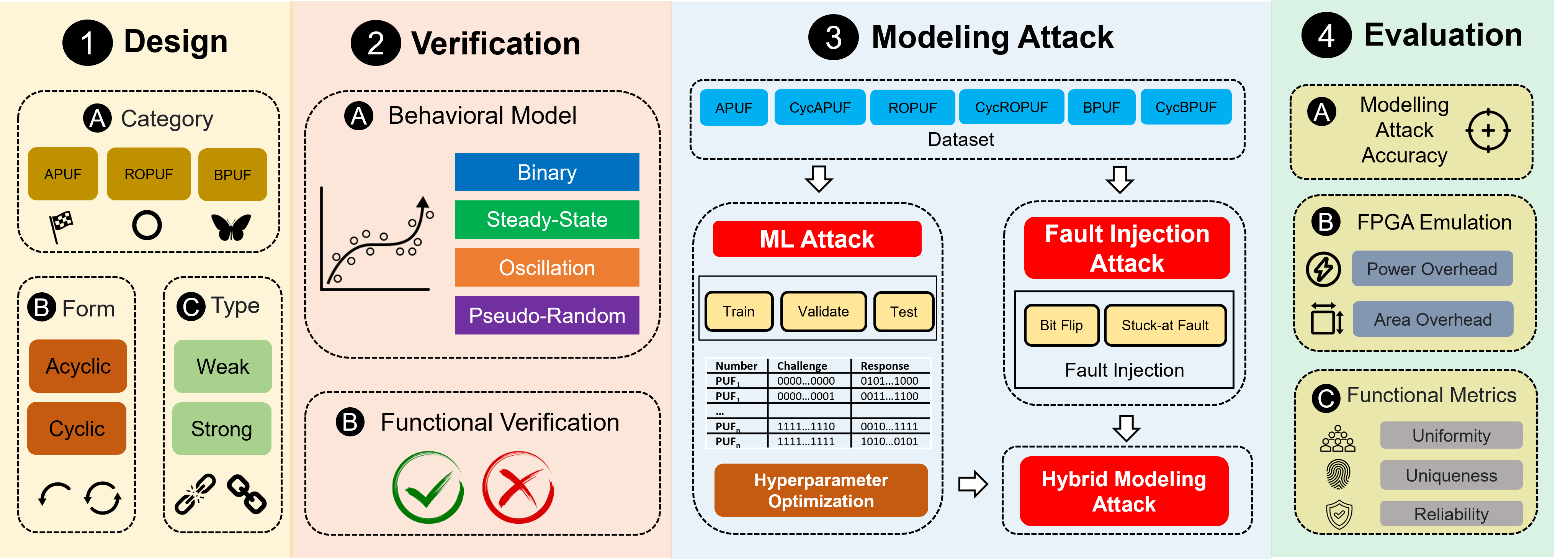}
    \caption{CycPUF framework}
    \label{fig:BigPicture}
\end{figure*}

\subsection{Related Works}
\label{Sec:Related}
There are various types of traditional delay-based PUFs, such as Arbiter PUF (APUF) \cite{Bhaaskaran2023}, Ring Oscillator PUF (ROPUF) \cite{Mahapatra2015}, and Butterfly PUF (BPUF) \cite{Tuyls2008}. Each type of PUF has its own unique properties and strengths, and the selection of the PUF category depends on the intended application and security requirements. For instance, ROPUFs rely on the frequency variations of multiple oscillators, while APUFs rely on the propagation delays of identical multiplexers \cite{Liu2019}. A unified PUF that generalizes various existing PUFs has also been modeled \cite{Sadeghi2016}.

One of the main attacks targeting the unpredictability of strong PUFs is the ML-based modeling attack, which has been shown to require only a linear or log-linear number of known CRPs to achieve high prediction accuracy on unknown CRPs \cite{Ruhrmair2010}. An attacker may use the constructed PUF model to attack the same PUF instance they collected the CRPs from, or they may also use a sophisticated enough modeling attack to attack multiple instances of a PUF design \cite{Naghmeh2022}. It is even demonstrated that it is feasible to model not just delay-based PUFs but memory-based memristor PUFs \cite{Sadeghi2013} with high prediction rates \cite{Sadeghi2020}. We have witnessed significant advancements in defense strategies aimed at bolstering the resilience of PUFs against ML-based attacks. However, as time progresses, we expect ML to become more and more sophisticated, meaning that down the line, these PUF designs may also fall victim to modeling susceptibility.

A deception authentication protocol has been proposed by deceiving the adversary to use a training set dominated by invalid responses \cite{Chongyan2021}. However, finding such a set for each instance of the device is a tedious task for the designer. While Interpose PUF (IPUF) \cite{Ruhrmair2018} has been proposed as an anti-modeling primitive replacement for APUF, it was successfully modeled using a divide-and-conquer-based attack methodology \cite{Ruhrmair2020}. In addition, structural unpredictability is exploited to reconfigure a conventional PUF into a noisy PUF and make it inherently unreliable and hence hard to model \cite{Lao2021}. The downside of such an approach is the need to look for reliable CRPs or embed an Error Correction Code (ECC) module to make some pre-selected CRPs reliable. Linear-Feedback Shift Registers (LFSR) have been utilized for the purpose of obfuscating the CRPs of the proposed PUFs \cite{Yao2021, Wanli2022}, as well as using auxiliary PUFs to accomplish a similar goal \cite{Karimi2021}. Further, to hide the direct relationship between CRPs, a dual-mode PUF is introduced using a feedback structure to perform bitwise {\sc xor} of the challenge and its response and use the result as input to challenge the PUF again \cite{Qu2018}. The method, however, seems to double the size of the original PUF. 

\subsection{Threat Model}
\label{Sec:Threat}
We assume that the attacker has physical access to the IC and can apply a polynomial number of challenges to the PUF module to collect the corresponding responses. With the measured CRPs of the PUF in hand, the adversary tries to build a numerical model of the PUF using ML algorithms. In addition, the adversary may leverage physical access and knowledge of the PUF design to carefully inject faults to alter the PUF's functionality. The introduced faults, such as bit flips or stuck-at faults, may cause the PUF to generate erroneous responses that deviate from its anticipated behavior.

\section{Preliminaries}
\label{Sec:Prelim}
The critical functional metrics needed to assess the functionality of a PUF are discussed here \cite{Maiti2013}. The Hamming Distance (HD) between two responses is defined as follows:
\begin{equation}
HD(R_1, R_2) = \sum_{i=1}^{n} (R_1[i] \oplus R_2[i])
\end{equation}
where $R_1$ and $R_2$ are two responses, and $n$ is the number of bits in the response vector.

In addition, the Hamming Weight (HW) of one response is the sum of existing `1's in the response vector, as follows:
\begin{equation}
    HW(R) = \sum_{i=1}^{n} R[i]
\end{equation}

\subsection{Uniqueness}
Different PUF instances of the same design shall ideally generate different responses under the same challenge. This is what separates one PUF from another. Uniqueness is the metric used to highlight the differences between different PUF instances, which can be defined as the normalized inter-chip HD as follows:
\begin{equation}
 \resizebox{0.91\hsize}{!}{
      $Uniqueness = (\frac{2}{k(k-1)} \sum_{i = 1}^{k-1} \sum_{j=i+1}^{k} \frac{HD(R_i, R_j)}{n}) \times 100\%$
      }
      \label{unique}
\end{equation}
in which for any pair of different PUF instances, PUF$_i$ and PUF$_j$, the responses produced by the involved PUFs are denoted by $R_i$ and $R_j$ respectively, and the number of PUF instances involved is given by the number $k$. The ideal uniqueness a PUF design can achieve is 50\%.

\subsection{Reliability}
\label{Sec:Reliable}
Ideally, PUFs would be able to operate in non-ideal conditions and still produce the same response for the same challenge that the PUF gets fed. Reliability is a metric used to identify a PUF's ability to reproduce the same response under the same challenge but under different operating conditions, such as different temperatures, altitudes, and supply voltages. Here, the intra-chip HD is used to measure reliability as follows:
\begin{equation}
    Reliability = (1 - \frac{1}{s} \sum_{i=1}^{s} \frac{HD(R, R^*_{i})}{n}) \times 100\%
    \label{reliable}
\end{equation}
where $R$ is the $n$-bit reference response from a single PUF operating in standard conditions, and $R^*_{i}$ is the response under the same challenge, collected under different operating conditions, and $s$ is the total number of samples collected from varying the conditions of operation. A truly reliable PUF will achieve a reliability of 100\%. This means that under varying operating conditions, the PUF can produce the same response under the same challenge.

\subsection{Uniformity}
\label{Sec:Uniform}
The uniformity of a PUF can be measured as the normalized HW of all the responses produced by the PUF. The formula for uniformity is given as follows:
\begin{equation}
    Uniformity = (\frac{1}{m}\sum_{i=1}^{m} \frac {{HW(R_i)}}{n}) \times 100\%
    \label{uniform}
\end{equation}
in which $m$ is the number of responses produced by the PUF. The ideal uniformity a PUF design can achieve is 50\%. 

\section{Cyclic PUF}
\label{Sec:CycPUF}
The proposed CycPUF framework is depicted in Fig. \ref{fig:BigPicture}. 

\circleNumber{1}While combinational feedback loops are generally not preferred when designing circuits, modifying acyclic PUF designs with feedback signals lends itself to interesting behaviors, which we shall refer to as \textit{response modes}, that can prove useful in hardware security. The main idea behind constructing a CycPUF is to take a few bits from the response vector of a delay-based PUF and route them back into the challenge vector. Specifically, we choose three traditional PUF designs (i.e., APUF, ROPUF, and BPUF), select some random response bits and then {\sc xor} each with a random challenge bit, and route the output of the {\sc xor} gates back into the challenge input of PUFs. 
This allows us to hold the challenge vector constant while the response vector remains free to change at will. In this case, the users can choose between different PUF categories (i.e., APUF, ROPUF, and BPUF), PUF form (i.e., acyclic or cyclic), and PUF type (i.e., weak or strong) based on their needs. 

\circleNumber{2}Based on our analysis, the output of CycPUF can behave in four different response modes under a constant challenge.

\begin{figure}[htb]
    \centering
     \includegraphics[width = \columnwidth]{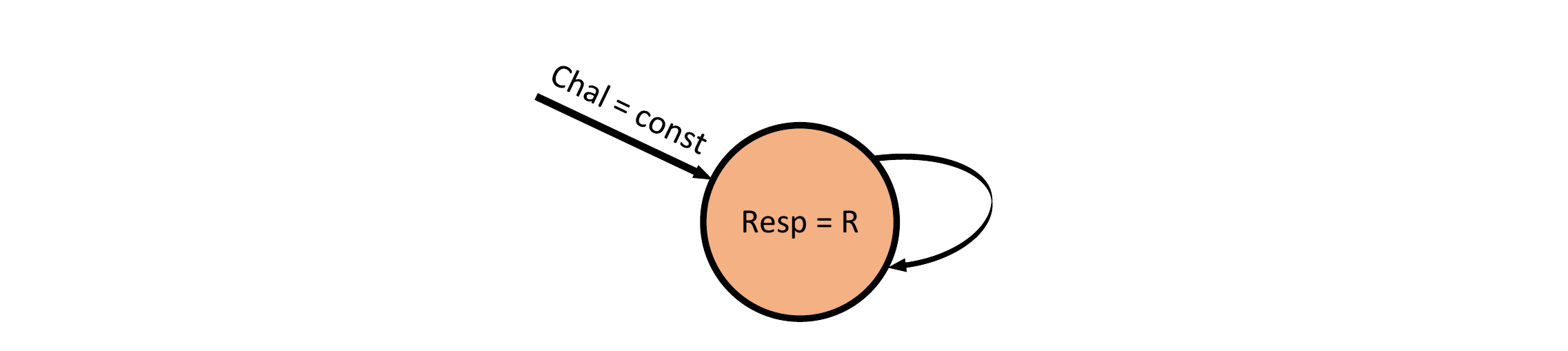}
    \caption{CycPUF binary response mode}
    \label{fig:binary}
\end{figure}

\textbf{1) Binary:} In the binary response mode (shown in Fig. \ref{fig:binary}) the CycPUF behaves the same as its acyclic counterpart. Under a fixed challenge, the CycPUF will generate a fixed response depending on what delay variations appear in their symmetric paths for the duration that this challenge is applied.

\begin{figure}[htb]
    \centering
     \includegraphics[width = \columnwidth, left]{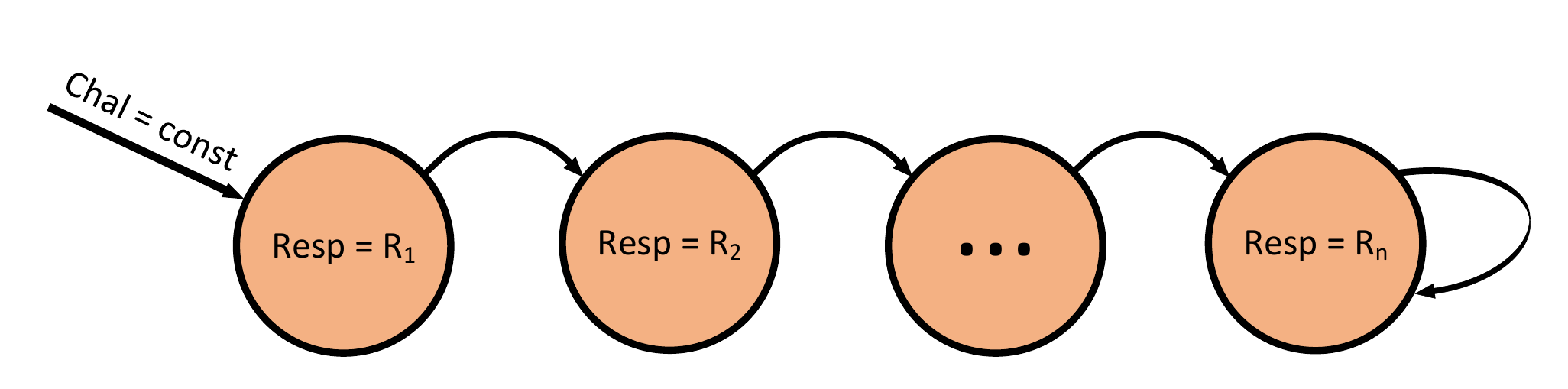}
    \caption{CycPUF steady-state response mode}
    \label{fig:steady}
\end{figure}

\textbf{2) Steady-State:} In the steady-state response mode (shown in Fig. \ref{fig:steady}), we see that inputting a fixed challenge will eventually produce a fixed response for the duration that the challenge is held constant. However, the fixed response may take some time to stabilize. This is to say that for some time, the CycPUF, while in this response mode, may produce different outputs before finally settling on a fixed response. This \textit{cool-down} period distinguishes this behavior from a PUF's usual acyclic behavior since, in this scenario, the CycPUF response needs no additional helper code or hardware to maintain its steady-state response mode.

\begin{figure}[htb]
    \centering
     \includegraphics[width = \columnwidth, left]{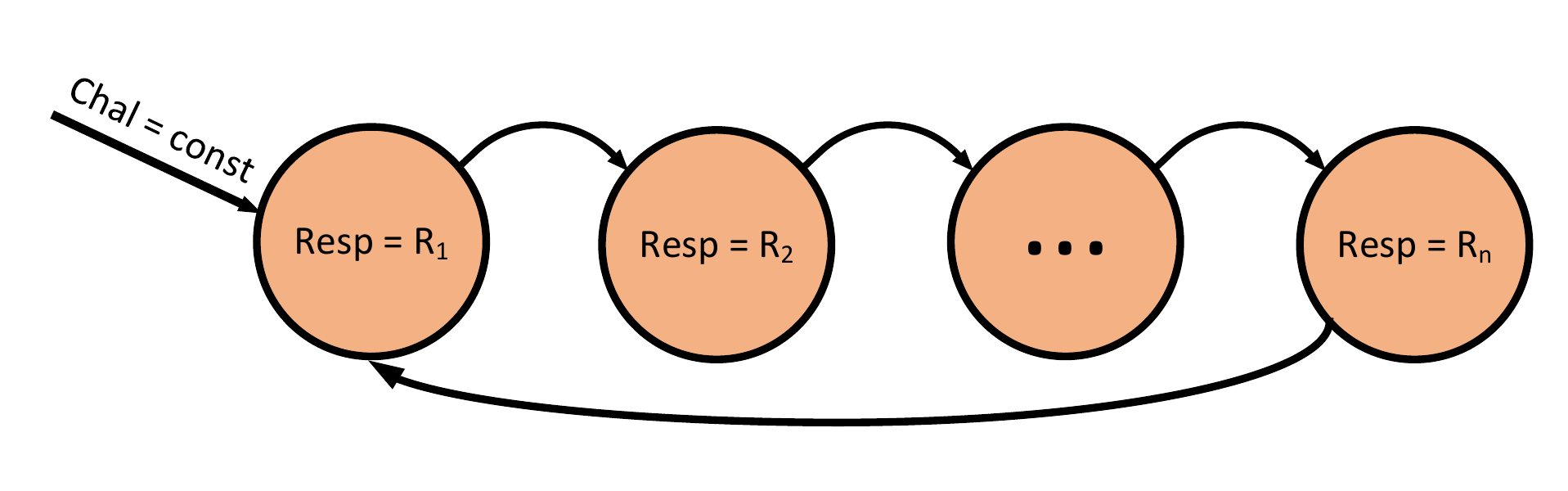}
    \caption{CycPUF oscillating response mode}
    \label{fig:oscillating}
\end{figure}

\textbf{3) Oscillating:} Oscillating outputs are also possible when using CycPUF (shown in Fig. \ref{fig:oscillating}). This oscillating response output can have varying periods for its output. It may take some intermediate responses before going back to a previously seen response. Following the same trend as the steady-state response mode's situation, we may also see a cool-down period before any oscillating outputs can be observed. When a fixed challenge is applied, the CycPUF may begin to produce its oscillating output after some time.

\begin{figure}[htb]
    \centering
     \includegraphics[width = \columnwidth, left]{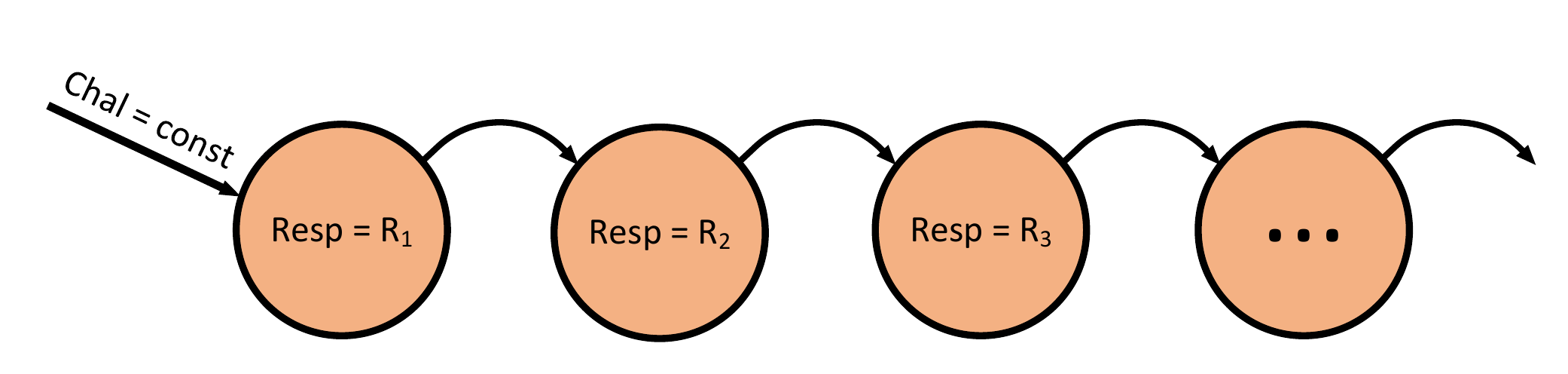}
    \caption{CycPUF pseudo-random response mode}
    \label{fig:random}
\end{figure}

\textbf{4) Pseudo-Random:} The pseudo-random response mode is the one where applying a fixed challenge to the CycPUF produces response vectors with no discernible pattern (shown in Fig. \ref{fig:random}). The steady-state and oscillating response modes may also produce seemingly random response vectors before finally collapsing to their appropriate behavior. However, the pseudo-random response mode does not follow a specific pattern, and it continuously generates separate responses for the duration of time that the challenge is held constant.

\begin{figure*}
    \centering
    \subfloat[]
    {\includegraphics[scale = .29]{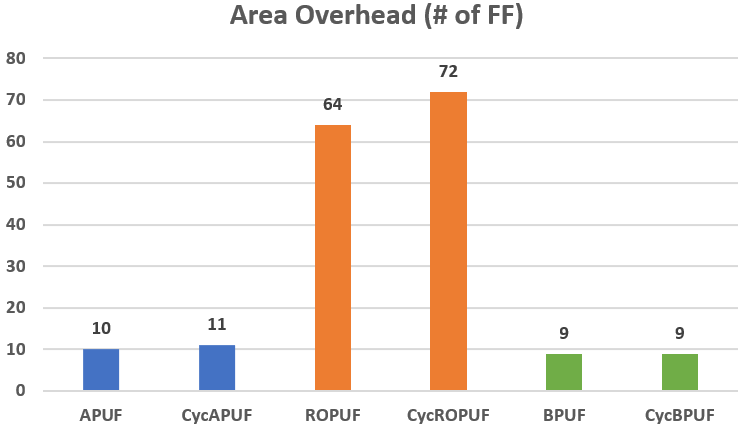}} 
    \hspace{1em} 
    \subfloat[]
    {\includegraphics[scale = .29]{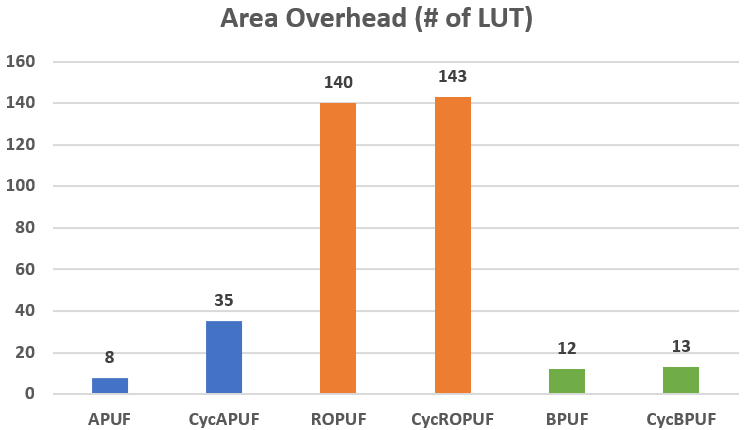}}
    \hspace{1em}
    \subfloat[]
    {\includegraphics[scale = .29]{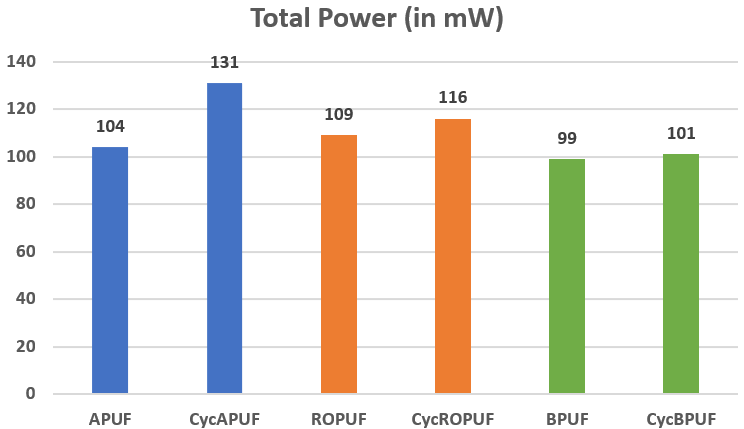}}
    \hspace{1em}
    \caption{Overhead analysis of CycPUFs and acyclic PUFs with 4-bit challenge and 4-bit response sizes (a) Number of FFs (b) Number of LUTs (c) Total power consumption}%
    \label{fig:overhead}
\end{figure*}

By cascading the delays of a PUF, the PUF is allowed to take on new possibilities that were otherwise thought to be unreliable. This gives rise to the term Challenge-Response Modes (CRMs), where instead of pairing up a single challenge with a single response, we now match a single challenge to the set of response vectors that the constant challenge may produce, where they will follow the behavior outlined in the given mode. Pairing a challenge in a CRM with each response vector in the set creates CRP-equivalents, where it should be noted that the challenges across these CRPs will overlap with each other. 

Take the steady-state response mode, for example. In this CRM, we see that the output produces a response pattern to a fixed challenge rather than a single response vector. For modeling attacks that expect an acyclic, reliable response from a PUF, the introduction of the steady-state as a CRM would be able to poison the data that a ML algorithm can gather and render the attack unsuccessful. Take the oscillating response mode as another example. Since specific patterns will be repeated, this CRM can be a potential case for device authentication in a way that checking only a single challenge and storing the sequences of responses allows us to validate the authenticity of an IC later on. Last but not least, the pseudo-random CRM can be utilized to generate pseudo-random IC-specific keys by checking only one challenge vector.

\circleNumber{3} After designing and verifying acyclic and cyclic PUFs, we apply a ML-based model-building attack \cite{Chakraborty2019} to test the security of the traditional PUFs and CycPUFs. While the attack is primarily designed for APUFs, it can be easily extended to other delay-based PUFs and, in our case, with some modification, to CycPUFs since the training set is based on a polynomial number of observed CRPs. In addition, we mimic the behavior of the fault attack on PUFs \cite{Tajik2015} and insert stuck-at and bit-flip faults into the PUFs, create a faulty dataset, and run a hybrid modeling attack.

\circleNumber{4} Finally, we evaluate the proposed CycPUF framework by reporting the accuracy of the modeling attack and then emulating the PUFs' behavior on FPGA boards. It helps us practically measure the power and area overheads as well as extract the mentioned functional metrics in Section \ref{Sec:Prelim}.

\section{Experimental Results}
\label{Sec:Experiments}
We wrote a Python script to generate cyclic APUF, ROPUF, and BPUF in Verilog format based on the user-input challenge and response size and the number of feedback signals. Then, we created different CycPUFs and validated their Register Transfer Level (RTL) and post-implementation behavior via testbenches. After that, we evaluated the overhead, security, and functional metrics for the created CycPUFs and compared them with their acyclic counterparts. For running simulations, we used the Xilinx Vivado HL WebPack tool version 16.4 and implemented the design on the Nexys A7 FPGA board.
{\renewcommand{\arraystretch}{1}%
\begin{table}[!t] 
    \centering
    \caption{Modeling attack results}
    \begin{tabular}{>{\centering\arraybackslash}p{.85in}>{\centering\arraybackslash}p{.45in}>{\centering\arraybackslash}p{.75in}>{\centering\arraybackslash}p{.75in}}
    \hline
        \textbf{PUF Design} & \textbf{Challenge Size} & \textbf{\# of Training CRPs} & \textbf{Model Accuracy} \\ \hline \hline
        APUF & 64 & 500,000 & 99.38\% \\ 
        CycAPUF & 64 & 703,340 & 59.49\% \\ 
        Faulty CycAPUF & 64 & 500,000 & 76.27\% \\
       \hline
        ROPUF & 64 & 500,000 & 78.00\% \\ 
        CycROPUF & 64 & 1,048,576 & 48.74\% \\ 
        Faulty CycROPUF & 64 & 642,603 & 50.02\% \\
        \hline
        BPUF & 64 & 500,000 & 83.32\% \\ 
        CycBPUF & 64 & 938,420 & 54.76\% \\
        Faulty CycBPUF & 64 & 582,645 & 61.44\% \\ \hline
    \end{tabular}
    \label{tab:attack}
\end{table}
}

\subsection{Security Evaluation}
\label{Sec:Security}
For security evaluation, we generated different strong PUFs with 64-bit challenge vectors and single-bit responses. This was done for compatibility between the PUF designs and the ML-based modeling attack used \cite{Chakraborty2019}. Each CycPUF design had a different number of feedback paths: 4 for the CycAPUF, 16 for the CycROPUF, and 12 for the CycBPUF. The CRP set was divided into two parts: the training set consisted of 80\% of the total CRPs, while the test set consisted of the remaining 20\%. We also ran a fault-injection modeling attack \cite{Tajik2015} by injecting a mix of stuck-at-0, stuck-at-1, and bit flip faults into the CycPUF designs and rerunning the ML-based modeling attack on the faulty CycPUFs. We injected 2 faults into the CycAPUF, 11 into the CycROPUF, and 7 into the CycBPUF. For the number of training CRPs, the acyclic PUFs all have 500,000, while the CycPUFs and faulty CycPUFs may possess a higher number of CRP-equivalents since, due to the cyclic behavior of the CycPUFs, they may generate multiple responses despite the challenge vectors being held constant.

The attack results are shown in Table \ref{tab:attack}. Please note that the higher the model accuracy, the more successful the attack is at modeling the PUF. Comparing the modeling accuracy of the acyclic PUFs with the CycPUFs, we see that there are clear improvements in resistance against modeling attacks by CycPUFs. The CycAPUF and APUF had the most impressive discrepancies between each other, producing a difference in modeling accuracy of 40\%. The CycROPUF and ROPUF had a model accuracy difference of 30\%, while the CycBPUF and BPUF ended with a model accuracy difference of 28.5\%. While injecting faults and running a hybrid attack can improve the modeling accuracy of the ML-based attack on CycPUFs, they still show better resistance compared with non-faulty acyclic ones. 

\subsection{Overhead Measurement}
\label{Sec:Overhead}
We generated weak cyclic and acyclic PUFs with 4-bit challenge and 4-bit response sizes and implemented them on the Nexys A7 FPGA board. The post-implementation overhead results are shown in Fig. \ref{fig:overhead}. 

For area overhead, we calculated the number of Look-Up Tables (LUTs) and Flip-Flops (FFs) utilized in the FPGA for each PUF shown in Fig. \ref{fig:overhead}a and Fig. \ref{fig:overhead}b. It can be seen that CycAPUF exhibits significantly higher hardware complexity, with an approximate 3x increase in the number of LUTs and a 10\% increase in the number of FFs compared to APUF. CycROPUF in contrast, approximately shows a 2\% increase in LUTs and a 12.5\% increase in FFs compared to ROPUF. CycBPUF maintains a relatively modest hardware overhead with an 8.3\% increase in LUTs and no increase in FFs. In addition, CycPUFs tend to exhibit slightly higher power consumption compared to acyclic PUFs. As can be seen in Fig. \ref{fig:overhead}c , the power consumption increase is 26\%, 6.5\% and 2\% for CycAPUF, CycROPUF, and CycBPUF compared with their acyclic counterparts. Further, under the same challenge and response sizes, BPUF and CycBPUF generally consume less power than the others. The selection of a PUF design should carefully consider hardware overhead and power consumption trade-offs in relation to the specific constraints and requirements of the intended application.  

\subsection{Functional Metrics Analysis}
\label{Sec:Functional}
Referring back to CycPUFs output behavior of Figs. \ref{fig:steady}, \ref{fig:oscillating}, and \ref{fig:random} in Section \ref{Sec:CycPUF}, trying to apply the outlined functional metrics of Formulas \ref{unique}, \ref{reliable}, and \ref{uniform} in Section \ref{Sec:Prelim} would not be immediately possible since under a fix challenge, there may be a set of responses rather than one response for CycPUFs. Instead, we shall define a modification to these metrics so that we are able to compare the proposed CycPUFs to their acyclic counterparts in a meaningful way. We know that under a constant challenge, any response bit in the response vector can be at a logic `1' or a logic `0'. Averaging the number of cycles for which each bit is high or low allows us to apply the functional metrics to get an insight into how the CycPUFs will perform.

Let $c$ be the number of clock cycles that a challenge is applied for, $R^i: i \in \{1, 2,..., c\}$ be a response vector among the set of response vectors that appear during that time, and $r^i_j \in R^i: j \in\{1, 2,..., n\}$ be a response bit in response vector $R^i$. Then, the Average Bit Value ($ABV$) can be defined as follows:
\begin{equation}
    ABV(j) = \frac{1}{c} \sum_{i=1}^{c} r^i_j
\end{equation}
If the resulting bit is greater than or equal to 0.5, then we say that on average we will observe a logic `1' from the output. Likewise, if the resulting bit is less than 0.5, we will observe a logic `0'. From here, we can apply $ABV$ to the three previously defined PUF metrics in Section \ref{Sec:Prelim}.

The PUF functional metrics were computed and compared for the acyclic and cyclic PUFs using the same setup as in Section \ref{Sec:Overhead}. The results are shown in Table \ref{tab:metrics}. As can be seen, while CycPUFs in general outperform their acyclic counterparts in uniqueness, they inherit the reasonable uniformity and reliability of their acyclic versions. As other papers have also previously highlighted \cite{Chakraborty2019, Maiti2013, Nguyen2018}, we confirmed that the FPGA-based APUF implementation exhibits relatively poor uniqueness. The intriguing fact is that CycAPUF fixes this issue and raises uniqueness to practically perfect levels. Comparing Tables \ref{tab:attack} and \ref{tab:metrics}, it can be observed that improvement in the uniqueness metric has a positive effect on improving resistance against modeling attacks.

{\renewcommand{\arraystretch}{1}%
\begin{table}[!t]
    \centering
    \caption{PUF functional metrics results}
    \begin{tabular}{>{\centering\arraybackslash}p{.85in}>{\centering\arraybackslash}p{.6in}>{\centering\arraybackslash}p{.6in}>{\centering\arraybackslash}p{.6in}}
    \hline
         \textbf{PUF Design} & \textbf{Uniqueness} & \textbf{Uniformity} & \textbf{Reliability} \\ \hline \hline
        APUF & 7.55\% & 55.42\% & 99.77\% \\
        CycAPUF & 47.10\% & 47.30\% & 98.34\% \\
        \hline
        ROPUF & 44.26\% & 53.81\% & 99.05\% \\
        CycROPUF & 50.68\% & 52.12\% & 95.18\% \\
        \hline
        BPUF & 11.48\% & 55.04\% & 97.45\% \\
        CycBPUF & 53.05\% & 46.67\% & 98.62\% \\
        \hline
    \end{tabular} 
    \label{tab:metrics}
\end{table}
}

\section{Conclusion}
\label{Sec:Conclusion}
In this paper, we introduced CycPUF, a novel lightweight PUF generation framework featuring strong resistance against ML-based and hybrid modeling attacks while keeping up the PUF functional metrics. CycPUFs use response feedback loops to introduce the notion of challenge-response modes and poison the data an adversary might use for model training, introducing inaccurate correlations by erratically adjusting the response vector in every cycle. This improves the functional metrics of conventional PUF designs at the same time. The attack results confirm that CycPUFs achieve a high level of unpredictability against ML-based and hybrid modeling attacks. Additionally, CycPUFs were able to attain near-ideal levels of uniqueness, uniformity, and reliability. This work offers researchers new perspectives on approaching hardware security problems with cyclic combinational circuits and looking into developing synthesis-friendly toolkits for them.

\section*{Acknowledgment}
This material is based upon work supported by the National Science Foundation under Award No. 2245247.

\bibliographystyle{IEEEtran}
\bibliography{IEEE}

\end{document}